\documentclass[prb,twocolumn,showpacs]{revtex4}
\usepackage{graphicx}
\usepackage{epsfig}
\usepackage{longtable}

\newcommand{\greeksym}[1]{{\usefont{U}{psy}{m}{n}#1}}

 \newcommand{\uGamma}{\mbox{\greeksym{G}}}

\begin{document}

\title{Electronic Structure of a Chain-like Compound: TlSe}
\author{\c{S}.~Ellialt{\i}o\u{g}lu,$^*$ E.~Mete,$^\dag$ R.~Shaltaf}
\affiliation{Department of Physics, Middle East Technical University, Ankara
06531, Turkey}
\author{and \\ K. Allakhverdiev,$^{1,2}$ F. Gashimzade,$^1$ M. Nizametdinova,$^3$ G.
Orudzhev$^3$} \affiliation{$^1$Institute of Physics, Azerbaijan National
Academy of Sciences, Baku 370073, Azerbaijan \\ $^2$Materials and Chemicals
Technologies Research Institute, MAM, T{\"U}B\.{I}TAK, Gebze/Kocaeli, Turkey \\
$^3$Azerbaijan Technical University, Baku, Azerbaijan}

\date{\today}

\begin{abstract}

An ab-initio pseudopotential calculation using density functional theory within
the local density approximation has been performed to investigate the
electronic properties of TlSe which is of chain-like crystal geometry. The
energy bands and effective masses along high symmetry directions, the density
of states and valence charge density distributions cut through various planes
are presented. The results have been discussed in terms of previously existing
experimental and theoretical data, and comparisons with similar compounds have
been made.

\end{abstract}

\maketitle

\section{Introduction}

\begin{figure}[htb]
\includegraphics[width=7.7cm,clip=true]{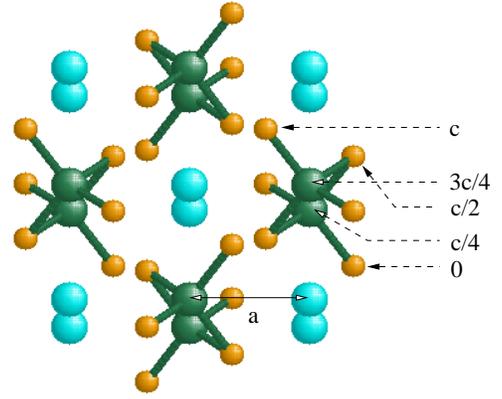}
\caption{Crystal structure of TlSe. Thallium atoms are represented by larger
spheres and selenium atoms by smaller ones in order to emphasize the chains for
better visualization. Se$^{2-}$ ions are tetrahedrally bonded to Tl$^{3+}$ ions
forming chains of edge-sharing-tetrahedra, along $c$-axis. The Tl$^+$ ions are
the spheres with no bonds shown.\label{figure1}}
\end{figure}

Thallium selenide is a III--VI compound with body-centered-tetragonal
(\emph{bct}) structure of D$^{18}_{4h}$ ($I4/mcm$) space group.\cite{kete} The
quasi-one-dimensional nature\cite{wyck} of its structure makes TlSe a generic
example for a series of similar binary and ternary chain-like compounds with
the formula unit of Tl$^{+}$(Tl$^{3+}$Se$^{2-}_2)^-$ where for the ternary
compounds monovalent and trivalent cations are of different atoms like in
TlInSe$_2$, TlGaTe$_2$, and TlInTe$_2$. The trivalent Tl$^{3+}$ ions are
surrounded by four tetrahedrally bonded Se$^{2-}$ ions. These tetrahedra share
edges to form long negatively charged chains of (Tl$^{3+}$Se$^{2-}_2)^-$ units
that are parallel to $z$-axis coinciding with the optical $c$-axis (see
Fig.~\ref{figure1}). Monovalent Tl$^+$ ions on the other hand are surrounded by
eight octahedrally positioned Se$^{2-}$ ions, and they electrostatically hold
these chains together by means of ionic inter-chain forces that are weaker than
the intra-chain bonds of Tl$^{3+}$--Se$^{2-}$ which are ionic-covalent in
nature. This leads to easy cleavage of TlSe-type crystals into plane parallel
mirror-like plates along $c$-axis. As a result one has a ``natural" (110) plane
with mirror-like surface that is particularly useful for optical measurements
and device applications. Also for the same reasons these crystals are highly
anisotropic in many physical properties which consequently are enhanced under
the influence of high pressures.\cite{GO81b,AMAE,AE} In addition, the
spin-exchange coupling between Tl$^+$ and Tl$^{3+}$ ions is observed to be
stronger\cite{PG} as compared to the intra-chain couplings of the same nature
between Tl$^{n+}$ and Tl$^{n+}$ ions.

The unit cell contains 8 atoms with atomic positions of two Tl$^{3+}$ ions at
$\pm$(0, 0, $c/4$), two Tl$^+$ ions at $\pm(a/2$, 0, $c/4)$ and four Se$^{2-}$
ions at $\pm(\eta a/2, \eta a/2, c/2)$ and $\pm(\eta a/2, -\eta a/2, 0)$, where
$\eta$ is the internal parameter. The primitive translation vectors are
($-a/2$, $a/2$, $c/2$), ($a/2$, $-a/2$, $c/2$) and ($a/2$, $a/2$, $-c/2$). The
lattice constants for TlSe can be found in the literature\cite{kete,wyck} as
$a$=$b$= 8.02$\pm$0.01 {\AA} and $c$=7.00$\pm$0.02 {\AA} with the internal
parameter $\eta$= 0.358.

The covalent bond between trivalent thallium and the divalent selenium is of
length d(Tl$^{3+}$--Se$^{2-}$)= 2.68 {\AA}, being just a little larger than the
sum of the covalent radii, 1.48 {\AA} and 1.16 {\AA}, respectively. The length
of ionic bond between the monovalent thallium placed in an octahedron of
divalent selenium ions is d(Tl$^+$--Se$^{2-}$)= 3.428 {\AA}, which is smaller
than but close to the sum of the respective ionic radii\cite{Shannon} of 1.59
{\AA} and 1.98 {\AA}. Other typical bond lengths are d(Tl$^{n+}$--Tl$^{n+}$)=
3.5 {\AA}, d$_1$(Se$^{2-}$--Se$^{2-}$)= 3.853 {\AA}, d(Tl$^{3+}$--Tl$^+$)= 4.01
{\AA} and d$_2$(Se$^{2-}$--Se$^{2-}$)= 4.06 {\AA}.

It has been shown that TlSe-type crystals are promising materials in device
applications as near- and far-infrared sensors, pressure sensitive
detectors\cite{KRA} and as $\gamma$--ray detectors.\cite{abdi} Switching
phenomena\cite{abay} and low-temperature metallic conductivity\cite{abdu92} in
TlSe have been described successfully.

The energy band gap published by different authors are not in accordance and
varies between 0.6 to 1.0 eV at 300 K.\cite{moos,IK} Band structure
calculations and comparisons with the existing experimental data showed that
TlSe-type materials are indirect gap materials; and the direct transitions are
forbidden according to the symmetry selection rules.\cite{IK} Electronic band
structures of TlSe and related crystals were obtained by various groups and
available in the literature.\cite{IK,GO80,GO81,JBNOS,KNM,okaz,orud}

Thallium selenide-type crystals posses a three dimensional electronic nature in
spite of their chain-like crystal structure. But still the direction normal to
the chains shows stronger band dispersion and consequently may be more
conductive. This can be seen from the experimental values\cite{IK} of direct
and indirect gaps for different polarization directions. The direct gap is
measured as 0.99 eV when the polarizations of the electric-field vector of the
incident electromagnetic wave \textbf{\emph{E}} is normal to the optical
$c$-axis ($z$-axis) whereas when \textbf{\emph{E}} is parallel to $c$-axis it
is measured to be 1.05 eV. Similarly, the indirect gap values are observed to
be 0.68 eV and 0.72 eV, respectively for the normal and parallel fields.

Effect of pressure and temperature on the electronic band structure of TlSe was
first reported by Gashimzade and Orudzhev,\cite{GO81b} and phase transitions in
TlSe-type crystals under pressure were reviewed by Allakhverdiev and
Ellialt{\i}o\u{g}lu.\cite{AE,AMAE}

In the present work the results of an \emph{ab-initio} pseudopotential
calculations using density functional theory within the local density
approximation for the electronic band structure as well as the density of
states of TlSe are presented. In addition, the valence charge density
distributions for various atomic planes are calculated. Effective masses for
various valleys in different symmetry directions were estimated by using
curvature fit to the bands. All these results are compared with the
experimental and other theoretical values available.

\section{Method}

We have used a pseudopotential method based on density functional theory in the
local density approximation. The self-consistent norm-conserving
pseudopotentials are generated by using the Troullier-Martins scheme \cite{TM}
which is included in the fhi98PP package.\cite{FS} Plane waves are used as a
basis set for the electronic wave functions. In order to solve the Kohn-Sham
equations,\cite{KS} conjugate gradients minimization method\cite{payne} is
employed as implemented by the ABINIT code.\cite{gonze} The
exchange-correlation effects are considered using the Perdew-Wang
scheme\cite{PW} as parametrized by Ceperley and Alder.\cite{CA}

Pseudopotentials are generated using the following electronic configurations:
for Tl, in addition to the true valence states (6$s$ and 6$p$), 5$d$ semicore
states are also included in the calculation. For Se, 4$s$ semicore and 4$p$
true valence states are treated as valence states. The optimized calculation
has produced the lattice parameters to be $a$=$b$= 7.91 {\AA} and $c$= 6.90
{\AA}, both of which are close to their experimental values within $\sim$1.4\%.

Good convergence has been obtained for the bulk total energy calculation with
the choice of a kinetic energy cut-off at 20 Ha for TlSe. In the density of
states calculations the irreducible Brillouin zone (BZ) was sampled with 80
${\bf \it k}$-points using the Monkhorst-Pack\cite{MP} scheme.

\begin{figure}[h]
\includegraphics[width=8.2cm,clip=true]{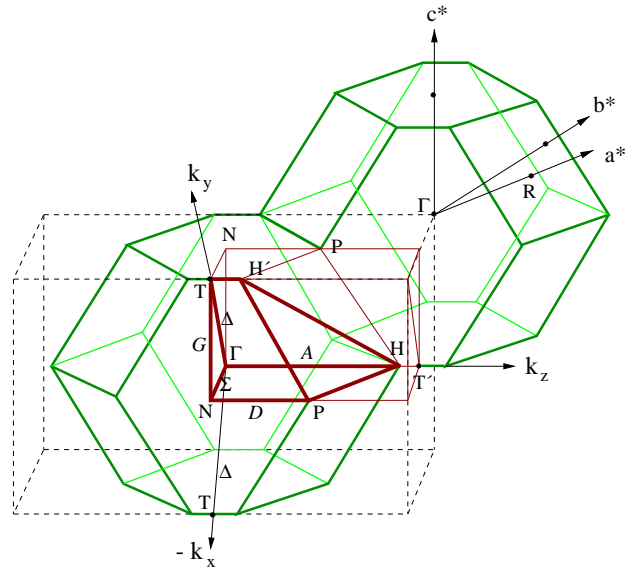}
\caption{The irreducible wedge (heavy lines) of the first Brillouin zone for
TlSe structure with the high symmetry points and high symmetry lines
indicated.\label{figure2}}
\end{figure}

\section{Results and Discussion}

\begin{figure*}
\includegraphics{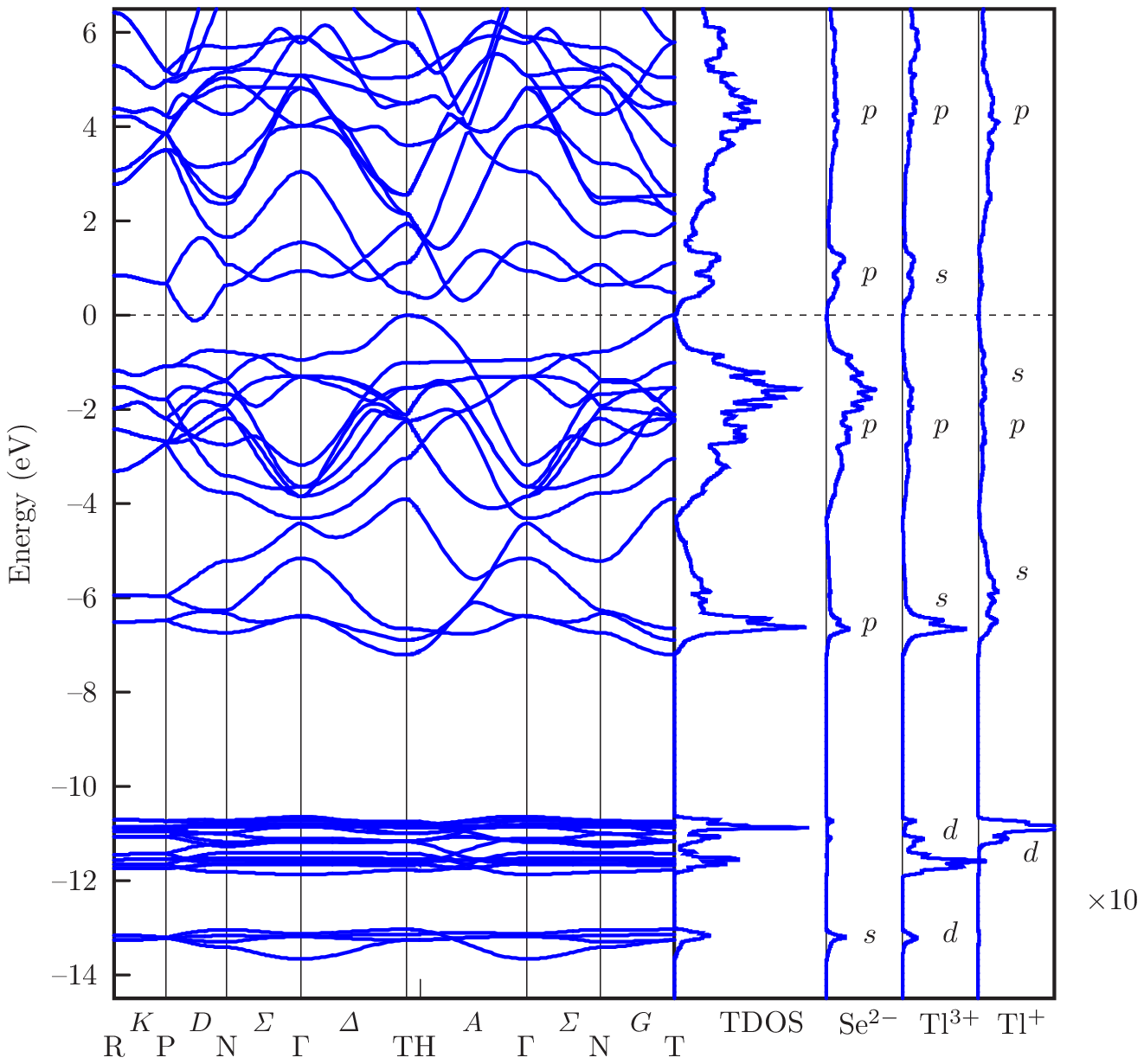}
\caption{The energy bands for TlSe along the high symmetry lines of the
Brillouin zone, the corresponding total density of states, and the local
densities of states for the Se$^{2-}$, Tl$^{3+}$, and Tl$^+$, in panels from
left to right, respectively. The densities of states for the lower valence
(semi-core) states due to Tl 5$d$ and Se 4$s$ electrons are shown in a scale
ten times smaller than the rest of the densities above --9 eV. The top of the
valence band is taken to be zero.\label{figure3}}
\end{figure*}

The first Brillouin zone for TlSe in the bct structure is given in
Fig.~\ref{figure2} where the high symmetry points and high symmetry lines are
indicated on the irreducible part (1/16th) of the BZ and they are given by:\\
R=(0, $\pi/a$, $\pi/c$), P=($\pi/a$, $\pi/a$, $\pi/c$), N=($\pi/a$, $\pi/a$,
0), $\Gamma$=(0, 0, 0), T=(0, 2$\pi/a$, 0) $\equiv$ T$^\prime$=(0, 0, 2$\pi/c$)
and H=(0, 0, (1+$u^2)\pi/c$) $\equiv$ H$^\prime$=(0, $2\pi/a$, (1--$u^2)\pi/c$)
where $u=c/a$. Symmetry lines of the Brillouin zone: $K$=($k$, $\pi/a$,
$\pi/c$), $D$=($\pi/a$, $\pi/a$, $k$), $\it\Sigma$=($k$, $k$, 0),
$\it\Delta$=(0, $k$, 0), $A$=(0, 0, $k$) and $G$=($k$, 2$\pi/a$--$k$, 0).

The dashed box in Fig.~\ref{figure2} has edges of $\sqrt{2}(2\pi/a) \times
\sqrt{2}(2\pi/a) \times 2(2\pi/c)$, the corners of which are the centers
($\Gamma^\prime$) of neighboring Brillouin zones. Similarly, the point
H$^\prime$ is equivalent to H since it is the H-point of the neighboring BZ.
The reciprocal lattice vectors for our choice of primitive translation vectors
are given by {\bf G$_1$}=(0, $2\pi/a$, $2\pi/c$), {\bf G$_2$}=($2\pi/a$, 0,
$2\pi/c$) and {\bf G$_3$}=($2\pi/a$, $2\pi/a$, 0) along ${\bf a}^*$, ${\bf
b}^*$ and ${\bf c}^*$ axes, respectively.

The energy bands calculated for the {\bf k}-points along the high symmetry
lines are shown in Fig.~\ref{figure3}. Also shown in the four rightmost panels
are the total density of states for the TlSe compound, as well as the local
densities of states for the individual ions, Se$^{2-}$, Tl$^{3+}$, and
Tl$^{+}$, respectively.

At the bottom of the figure there are 4 bands originated from 4$s$ states of
the Se atoms located in the range 13 to 13.7 eV below the top of the valence
band, which is chosen as zero. Above this group there are 20 bands consisted of
5$d$ states of Tl atoms located in the range from --10.6 to --12 eV, and being
semicore $d$-states they are not much dispersed in most {\bf \em k}-directions,
except a little along $D$, and along $A$.

In the region between --4 and --7 eV, there is an isolated group of 4 bands
which are made up of mostly the 6$s$ states of monovalent Tl ion, and some Se
4$p$ states at the bottom of the valence band mixed with the 6$s$ states of
trivalent cation. Another group consisting 10 bands in the upper part of the
valence band is mainly due to Se 4$p$--states and 6$p$ states of both Tl atoms.
However, the uppermost valence band which tops at the symmetry point T is
composed of mainly nonbonding Se 4$p$ and 6$s$ states of monovalent Tl.

The 2 bands in the lower part of the conduction bands are the antibonding
mixture of the Se 4$p$-states with the 6$s$ states of Tl$^{3+}$, and intermixed
only slightly (around H) with the 12 upper conduction bands that are located
above $\sim$1.5 eV, and made up of $p$-states of all three ions.

All of the bands along the line $K$ joining R and P points are doubly
degenerate due to time reversal symmetry.\cite{GO81}

The bottom of the conduction band is located almost at the midway
D$_1$=($\pi/a$, $\pi/a$, $\pi/2c$), along the line $D$ joining the points P and
N, and corresponds to the irreducible representation D$_1$. Two additional
minima are situated along the symmetry line $A$ which connects the points
$\Gamma$ and H, one very close to the midway (A$_4$) and the other is very
close to the point H. The band at point T is a little higher in energy.

The top of the valence band is sharply defined and located at the high symmetry
point T which corresponds to the irreducible representation T$_3$. The
T$_3\rightarrow$T$_4$ vertical transition is forbidden in the dipole
approximation, however, minimal direct transition is allowed at point H (near T
along $A$). The bottom of the valence bands is also at the symmetry point T
with $s$-like minimum, and the valence band width is found to be 7.21 eV.

\begin{figure}[h]
\includegraphics[width=8.7cm,clip=true]{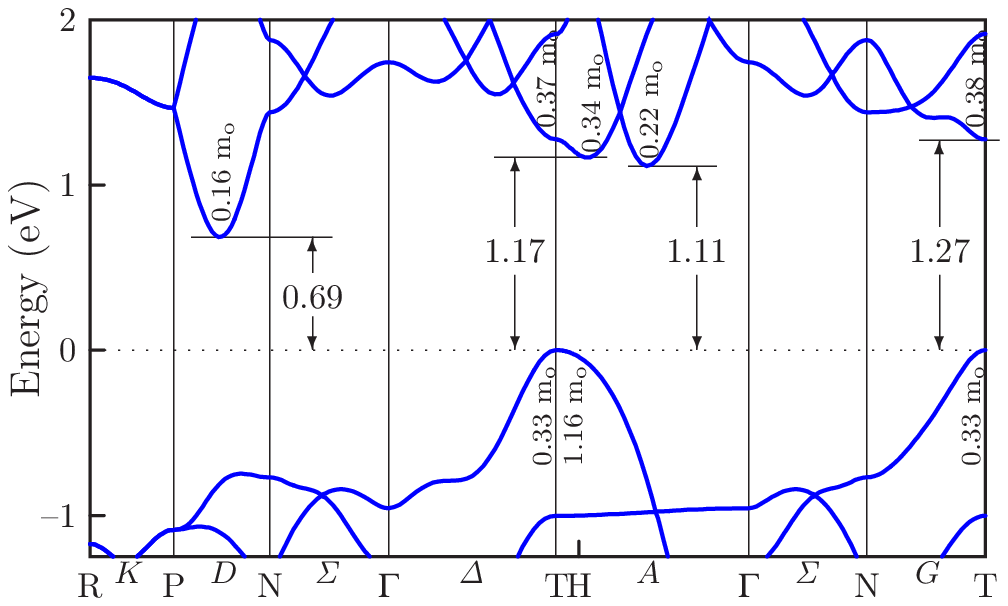}
\caption{Energy gap values and effective masses for valleys along various
directions. The conduction bands are rigidly shifted upward in energy by a
fixed amount of 0.8 eV (see the text)\label{figure4}}
\end{figure}

The energy gap is underestimated relative to the experimental value due to the
well known artifact of the LDA calculations,\cite{JG} as a result that the
indirect gap appears to be negative which instead should be about 0.7 eV. Note
that the similar deficiency is observed by Okazaki et al.\cite{okaz} for the
semiconductor TlGaTe$_2$ using ab-initio LAPW method, and resulting to a band
structure with slightly negative gap also, leading to a semimetal with hole
pocket at T$_3$ and electron pocket at D$_1$. Hence, for a rough correction, if
the whole conduction band (of 14 states) as a block is rigidly shifted upwards
by 0.8 eV to arrive at the indirect gap value (0.69 eV seen in
Fig.~\ref{figure4}) of TlSe then a direct gap of 1.27 eV is obtained as a
consequence, which is to be compared with the experimental value of roughly 1.0
eV.\cite{IK} The other indirect gaps will then be seen as 1.11 eV for the
valley along $A$ and 1.17 eV for the valley along the same direction but just a
few meV's away from H-point. The curvature of the lowest conduction band along
TH has the same sign but is larger than that of the highest valence band along
the same edge of the BZ, and thus the vertical gap at H-point being 1.22 eV is
0.05 eV smaller than the direct gap at T-point where the valence band top is
located.

From the curvatures of bands the effective masses are obtained by linear
fitting of E versus $k^2$ plots at the close proximity of the extrema. They are
shown in Fig.~\ref{figure4} for different valleys in units of free electron
mass $m_{\rm o}$. The hole effective masses along $\it\Delta$ and $\it\Sigma$
are found to be very close to each other being $m_h^*(\rm T_3)$=\,0.33\,$m_{\rm
o}$, however, along TH it is as high as $m_h^*(\rm T_3)$=\,1.16\,$m_{\rm o}$
and increasing curvature after H along $A$ reduces it to 0.61\,$m_{\rm o}$ (at
H) giving rise to a combined value of 0.81\,$m_{\rm o}$ at T. The electron
effective masses at T along $\it\Delta$ and $\it\Sigma$ are again similar and
have values of $m_e^*(\rm T_4)$=\,0.37\,$m_{\rm o}$ and 0.38\,$m_{\rm o}$,
respectively. Along TH the curvature of this band is opposite of those along
$\it\Delta$ and $G$ causing T$_4$ to be a saddle point as is well-known. Other
valleys along $A$ correspond to effective masses of 0.34\,$m_{\rm o}$ and
0.22\,$m_{\rm o}$ as seen in Fig.~\ref{figure4} and the electron effective mass
for the lowest minimum along $D$ is found to be 0.16\,$m_{\rm o}$. The
experimental values obtained from conductivity and Hall effect
measurements\cite{Gus} are $m_e^*$=\,0.3\,$m_{\rm o}$ and
$m_h^*$=\,0.6\,$m_{\rm o}$ and from thermoelectric measurements\cite{Ali} is
$m_h^*$=\,0.86\,$m_{\rm o}$.

\begin{figure}[h]
\includegraphics[width=7cm,clip=true]{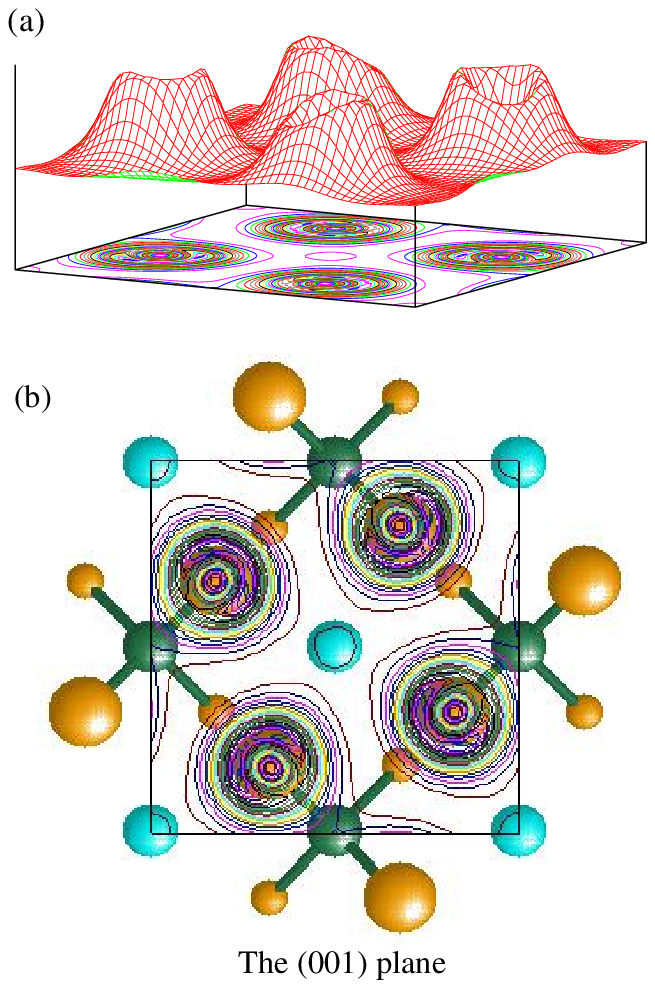}
\caption{(a) Oblique surface plots and (b) top views of the total valence
charge density contours for TlSe cut through the top (001) plane containing the
Se$^{2-}$ ions (larger balls). Tl ions (medium balls) and half of the Se ions
(smallest balls) are in (004) and (002) planes, respectively.\label{figure5}}
\end{figure}

\begin{figure}[ht]
\includegraphics[width=7cm,clip=true]{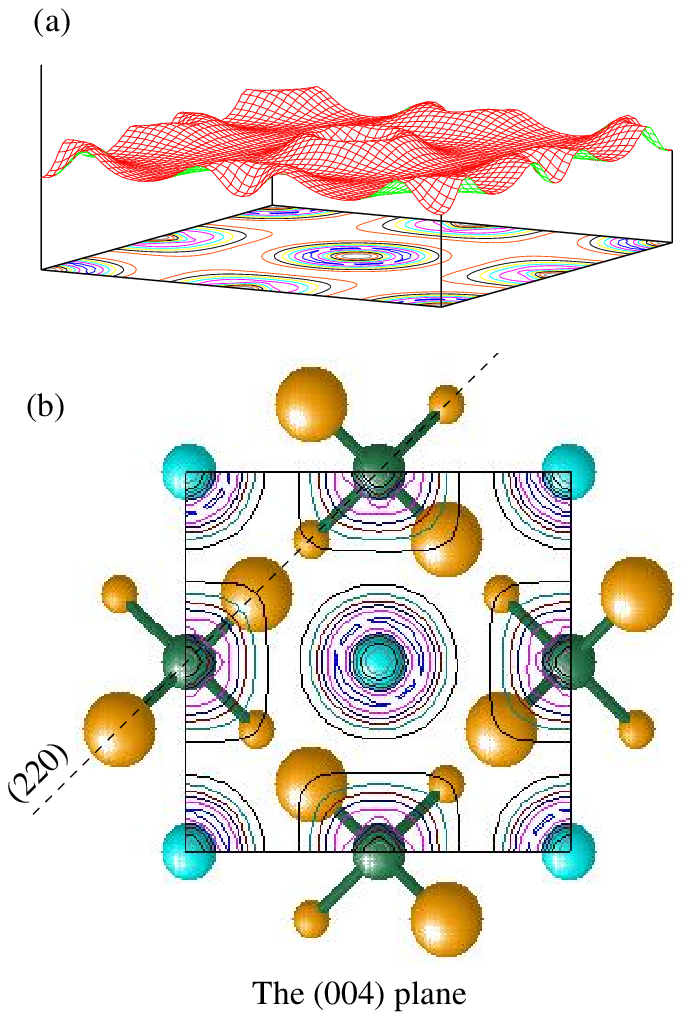}
\caption{(a) Oblique surface plots and (b) top view of the total valence charge
density contours for TlSe cut through the (004) plane containing both Tl$^+$
(center and corners) and Tl$^{3+}$ (edges) ions. Se$^{2-}$ ions shown are not
in plane, but either above (larger balls) or below (smaller balls) by $c$/4.
\label{figure6}}
\end{figure}

The total valence charge densities for different planes of atoms were
calculated to show the charge transfer which are in accordance with the local
density of states results in identifying the electronic structure of the
compound. Fig.~\ref{figure5} shows the plot for the (001) plane which cuts
through the Se$^{2-}$ ions at positions $\pm(\eta a/2, -\eta a/2, 0)$. The
bottom plane of (002) containing the other Se$^{2-}$ pair of the same
tetrahedron gives the same charge density plot, except it is flipped in $x$- or
$y$-axis.
\begin{figure}
\includegraphics[width=7cm,clip=true]{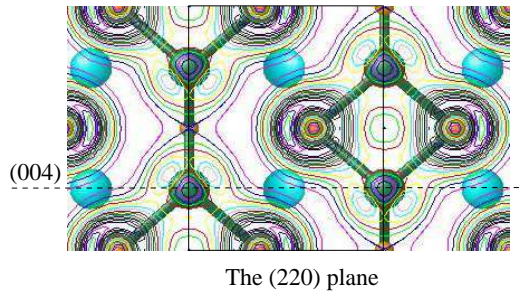}
\caption{Contour plots of the total valence charge density for TlSe cut through
the (220) plane containing Tl$^{3+}$ and Se$^{2-}$ ions. Tl$^{+}$ ions shown
are not in plane. The frame shows the part of (220) plane inside the cube shown
in Fig.~\ref{figure6} \label{figure7}}
\end{figure}

\begin{figure}[t]
\includegraphics[width=8cm,clip=true]{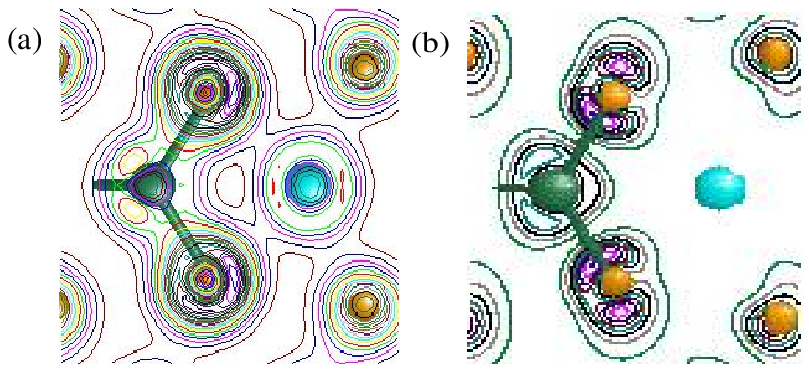}
\caption{Contour plots of, (a) the total valence charge density and (b) the
charge density due to the lowest two conduction bands, for TlSe cut through an
incommensurate plane containing all three ions. \label{figure8}}
\end{figure}

Fig.~\ref{figure6} shows the case for the (004) plane that contains Tl$^{3+}$
ions at $\pm$(0, 0, $c/4$) and Tl$^+$ ions at $\pm(a/2$, 0, $c/4)$. Tl$^+$ ions
having been stripped from their 6$p$ electron show their $s$-like character due
to the outermost 6$s^2$ electrons participating in the valence bands. This can
also be seen in the energy band picture as rather quadratic ($s$-like) minimum
and maximum at the high symmetry point T (see Fig.~\ref{figure3}). Donated
their 6$p$ and 6$s$ electrons to bond formation, Tl$^{3+}$ ions, on the other
hand, show some charge density extending towards Se$^{2-}$ ions and a
negligible amount of $d$-influence in Fig.~\ref{figure6}(b).

These findings can only be roughly compared with the earlier empirical
calculations\cite{OED} since the charge density contours for (001), (002) and
(004) planes were superimposed in one plot and the amplitude variations along
the bonds were depicted in a different figure. Therefore, in order to compare
the pronounced charge accumulation at the Tl$^{3+}$--Se$^{2-}$ bonds found in
the empirical calculation, charge density distribution on two other planes both
passing through these bonds are also presented. Fig.~\ref{figure7} shows the
total valence charge density plots for the (220) plane that passes through the
Tl$^{3+}$--Se$^{2-}$ bonds, where most of the charge is seen to be accumulated
on the Se$^{2-}$ ion rather than on the bond. The second plane that contains
the bonds under consideration contains also the monovalent cation. The total
valence charge density calculated for this plane is shown in
Fig.~\ref{figure8}(a), where the monovalent cation is seen to be not bonded to
the chalcogen ions. Again the Tl$^{3+}$--Se$^{2-}$ bond is seen to be more
ionic than covalent in nature. Finally, Fig.~\ref{figure8}(b) shows the same
plane that contains all three ions as (a), however, this time the charge
density contours depict the contributions from the lowest two conduction bands
only. It is clearly seen that the monovalent cation has no electron at these
antibonding bands which are formed by Tl$^{3+}$ 6$s$ and Se$^{2-}$ 4$p$ states
alone, consistent with the rightmost panel in Fig.~\ref{figure3}.

\section{Summary and Conclusion}

\emph{Ab-initio} pseudopotential calculations using density functional theory
within the local density approximation has been performed for the first time to
investigate the electronic properties of TlSe. The energy bands along different
symmetry directions, the local and total densities of states, and total valence
charge density distributions for certain plane-cuts are presented.

It is shown that the top of the valence band is sharply defined and located at
the high symmetry point T, which corresponds to the irreducible representation
T$_3$. The bottom of the conduction band is located almost at the midway
($\pi/a$, $\pi/a$, $\pi/2c$) between symmetry points P and N, along the the
line $D$, and corresponds to the irreducible representation D$_1$.

The distribution, dispersion and the orbital characters of the bands are in
good agreement with the experimental data of the photoemission.\cite{PT} The
only prior band structure calculation for TlSe is reported by Gashimzade et
al.\cite{GO81} where they have used a model pseudopotential and applied an
empirical method. Although there is a general agreement between the two
results, there are some major differences as well. In their result there is an
isolated group consisted of the top 2 valence bands which does not appear
separated from the rest of the valence band in our calculations. And the
isolated group of 4 bands at the bottom of the valence band in our calculations
is not separated from the rest of the valence band in their results. Moreover,
the order of their bands at the symmetry point N does not agree for most bands
with that of the present results. These differences arise from the fact that
Gashimzade et al.\cite{GO81} used restricted number of plane waves in their
empirical pseudopotential method, and moreover, the wave functions were not
deconvoluted enough. It is known that in such approximations it is not possible
to take into account the screening, and to include the electron-ion interaction
as well as the exchange-correlation effects properly. Another difference with
this study is, although they have included the 4$s$ semicore states of Se
atoms, the 5$d$ states of Tl atoms were not taken into account in their
calculations. To single out the effect of $d$ states, we have made separate
calculations with and without including the 5$d$ semicore states of Tl. Upon
inclusion of $d$ states the main change was on the 4$s$ semicore states of Se
which got narrowed down from dispersion between --13.7 eV to --12.2 eV to
--13.7 eV to --13 eV, decreasing the bandwidth from 1.4 eV to 0.7 eV. The other
change occurred in the lowest two conduction bands that are shifted downward
and separated from the rest of the conduction bands by about 0.2 eV in all
symmetry directions.

It is worth noting here that a similar electronic band structure for TlInSe$_2$
is obtained from a calculation\cite{orud} made by constructing the
pseudopotentials using the scheme suggested by Bachelet et al.\cite{BHS}
Although it is a semiconductor with a direct gap, valence band structure is
qualitatively analogous to that of TlSe presented here. Isolation of the group
of lowest 4 valence bands, the shape of the highest valence band, especially
along the $\uGamma$TH$\uGamma$ symmetry directions, and sharpness of the
valence band top are their similar behavior. Another material to be compared is
the TlGaTe$_2$, ab-initio band structure of which has been published by Okazaki
et al.\cite{okaz} Not mentioning the quantitative differences, in general, the
two band structures are remarkably similar, apart from some variations in the
vicinity of $\uGamma$ point of the Brillouin zone.

\begin{acknowledgments}
This work was supported by T{\"U}B\.{I}TAK, The Scientific and Technical
Research Council of Turkey, Grant No. TBAG-2036 (101T058).
\end{acknowledgments}
~\\[4mm]
\noindent $^*$ Corresponding author. Email: sinasi@metu.edu.tr \\[3mm]
\noindent $^\dag$ Permanent address: Bal{\i}kesir {\"U}niversitesi, Fizik
B{\"o}l{\"u}-m{\"u}, Bal{\i}kesir 10100, Turkey.
\bibliography{basename of .bib file}

\end{document}